\documentclass[12pt]{iopart}
\usepackage{graphicx}
\usepackage{amssymb}

\begin{document}

\title[Antiferromagnetic Order of Strongly Interaction Fermions in a Trap]{Antiferromagnetic Order of Strongly Interacting Fermions in a Trap: Real-Space Dynamical Mean-Field Analysis}

\author{M Snoek$^{1}$, I Titvinidze$^{1}$, C T\H oke$^{1}$, K Byczuk$^{2,3}$ and W Hofstetter$^{1}$}


\address{$^{1}$Institut f\"ur Theoretische Physik, Johann Wolfgang Goethe-Universit\"at, 60438 Frankfurt/Main, Germany} 
\address{$^{2}$Theoretical Physics III, Center for Electronic Correlations and Magnetism, Institute for Physics,
University of Augsburg, 86135 Augsburg, Germany }
\address{$^{3}$Institute of Theoretical Physics, Warsaw University, ul.~Ho\.za~69, 00-681 Warszawa, Poland}
\ead{snoek@itp.uni-frankfurt.de}

\date{\today}


\begin{abstract}

We apply Dynamical Mean-Field Theory to strongly interacting fermions in an inhomogeneous environment.
With the help of this Real-Space Dynamical Mean-Field Theory (R-DMFT) we investigate 
antiferromagnetic states of repulsively interacting fermions with spin $\frac{1}{2}$ in a harmonic potential.
Within R-DMFT, antiferromagnetic order is found to be stable in spatial regions with total particle density close to one, but persists also in parts of the system where the local density significantly deviates from half filling. 
In systems with spin imbalance, we find that antiferromagnetism is gradually suppressed and phase separation emerges beyond a critical value of the spin imbalance.

\end{abstract}

\submitto{\NJP}

\maketitle

\section{Introduction}
Ultracold atoms in optical lattices provide a new laboratory for interacting quantum many body systems \cite{Bloch}.
Bosonic atoms in optical lattices realize the Bose-Hubbard model \cite{Jaksch98} and undergo a superfluid-Mott insulating transition when the potential depth of the optical lattice is increased \cite{Greiner}. Recent experiments directly observed correlated particle tunneling \cite{Foelling07} and superexchange \cite{Trotzky08}, which are the basic mechanisms underlying quantum antiferromagnetism. Also the Fermi
surface of fermionic atoms in optical lattices \cite{Kohl} and fermionic superfluidity of attractively interacting lattice fermions \cite{Chin06} were recently observed, bringing the realization of strongly correlated many-fermion states nearby \cite{Hofstetter02}. A two-component mixture of repulsively interacting fermions (e.g. $^6$Li or $^{40}$K)
at half filling is predicted to form a correlated paramagnetic Mott insulator state above
the critical (N\'eel) temperature and antiferromagnetic order below this temperature \cite{georges07}. Reaching this temperature is predicted to be within today's experimental capabilities \cite{Werner05, Koetsier07}.

In contrast to solid state systems, lattice defects, impurities and phonons are absent in optical lattices. However, the spatial inhomogeneity due to
the harmonic confinement potential is always present, leading to a spatially varying local density.
Therefore, the concept of long range order is questionable and ordered phases are expected to develop on  finite length scales. 
A Hartree-Fock static mean-field theory predicts 
that antiferromagnetism, with staggered magnetization on a finite length scale, coexists with paramagnetic states in various spatial patterns, e.g. antiferromagnetism in the center of the trap or antiferromagnetism in a ring 
surrounded by a particle- or a hole-doped 
atomic liquid \cite{Andersen}.
On the other hand, both commensurate and incommensurate spin-density-waves have been predicted for the hole-doped Hubbard model \cite{incommensurate1, incommensurate2, incommensurate3}.
However, the existence and properties of any ordered state on a finite length scale are strongly sensitive to quantum and thermal fluctuations. 
Therefore a theoretical description that captures effects of strong correlations and spatial inhomogeneity in a unified framework is needed.
In this paper we apply a Real-Space Dynamical Mean-Field Theory (R-DMFT), 
which is a comprehensive, thermodynamically consistent and conserving  mean-field theory for correlated lattice fermions in the presence of an external inhomogeneous potential. The R-DMFT takes into account local correlations exactly \cite{Georges1, Georges2, Georges3, Georges4}. 

We prove that for spin-$\frac{1}{2}$ lattice fermions with local repulsive interaction antiferromagnetic order exists at zero temperature when the harmonic potential is present. We find that antiferromagnetic order is stable in spatial regions with total particle density close to one, but persists also in parts of the system where the local density significantly deviates from half filling. 
We also show that for strong repulsion phase separation occurs in imbalanced mixtures, when the difference in the particle-number of the spin components is large. For weaker repulsion a strong imbalance destroys the antiferromagnetic order, but does not lead to phase separation. These results are especially intriguing with respect to
recent experiments on attractively interacting fermions with spin imbalance, which have led to still unresolved questions regarding the nature of the observed phase separation \cite{Ketterle1, Ketterle2, Hulet}. 

\section{Model}
Repulsively interacting fermions in an optical lattice almost perfectly implement the Hubbard Hamiltonian
\begin{equation} \label{Hamiltonian}
\mathcal{H} =-J \hspace{-1mm} \sum_{\langle ij\rangle,\sigma} c^\dag_{i\sigma}c_{j\sigma} +U\sum_{i}n_{i\uparrow}n_{i\downarrow} + \sum_{i\sigma} (V_i - \mu_\sigma) n_{i\sigma},
\end{equation}
where
$n_{i\sigma}=c^\dag_{i\sigma}c_{i\sigma}$, and $c_{i\sigma}$ ($c^{\dagger}_{i\sigma}$) are fermionic annihilation (creation) operators for an atom with spin $\sigma$ at site $i$, 
$J$ is the hopping amplitude between nearest neighbor sites $\langle ij\rangle$,
$U>0$ is the on-site interaction, $\mu_\sigma$ is the (spin-dependent) chemical potential and $V_i=V_0 r_i^2$ is the harmonic confinement potential. Moreover we define $\bar \mu \equiv \frac{1}{2}(\mu_\uparrow+\mu_\downarrow)$ and $\Delta \mu \equiv \mu_\uparrow-\mu_\downarrow$.
The parameters of this model are tunable in experiments by a change of the lattice amplitude and via Fesh\-bach resonances \cite{Bloch}. In the following, $J=1$ sets the energy unit and we take the lattice constant to be  $a=1$. 

\begin{figure}
\begin{center}
\vspace{-1cm}
\includegraphics[scale = 0.75, keepaspectratio]{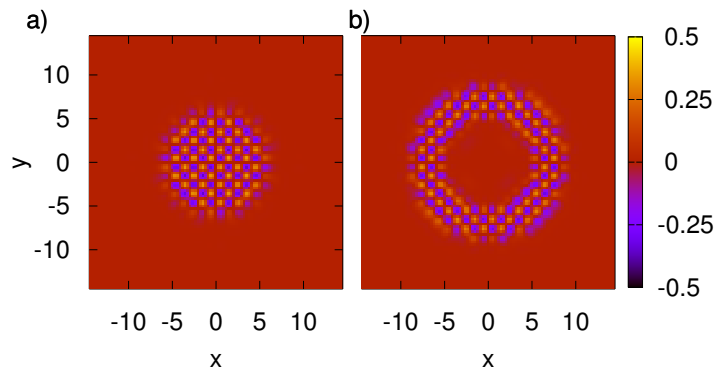}
\end{center}
\vspace{-1cm}
\caption{Real-space magnetization profiles for $U=10$ on a square (30$\times$30) lattice; a) $V=0.1$ and $\mu_\uparrow = \mu_\downarrow = 5$; b) $V=0.2$ and $\mu_\uparrow = \mu_\downarrow = 15$. Energies are expressed in units of the hopping parameter $J$.}
\label{pics}
\end{figure}

\section{Method}
To obtain the ground state properties of this system, we apply a real-space extension 
of dynamical mean-field theory (DMFT) \cite{Georges1, Georges2, Georges3, Georges4, Metzner1, Metzner2}.
Within R-DMFT the self-energy is taken to be local, which is exact in the infinite dimensional limit \cite{Metzner1, Metzner2}. However, in an inhomogeneous system it depends on the lattice site, i.e. $\Sigma_{ij\sigma}=\Sigma_{\sigma}^{(i)}\delta_{ij}$, where $\delta_{ij}$ is a Kronecker delta.
Formerly, a similar scheme has been developped for systems with inhomogeneity in one direction \cite{Pothoff99}. Only recently, models with full inhomogeneity have been investigated, in particular the Falicov-Kimball model \cite{Tran1, Tran2}, disorderded systems \cite{Song} and paramagnetic states of cold fermionic atoms \cite{Helmes}.

In the R-DMFT method, the Hamiltonian is mapped onto a set of single site problems. 
The physics of lattice site $i$ is described by the local effective action \cite{Georges3}
\begin{eqnarray}
\label{Seff}
S^{(i)}_{\rm eff}&=&-\int d\tau\int d\tau' \sum_{\sigma}c^\dag_{i\sigma}(\tau)\mathcal G^{(i)}_0(\sigma,\tau-\tau')^{-1}c_{i\sigma}(\tau')\nonumber\\
&&-U\int d\tau n_{i\uparrow}(\tau)n_{i\downarrow}(\tau),
\end{eqnarray}
which explicitly depend on the site index $i$. Here $\tau$ is imaginary time. The function $\mathcal G^{(i)}_0(\sigma,\tau-\tau')$ is a local non-interacting propagator interpreted as a dynamical Weiss mean-field which simulates the effect of all other sites \cite{Georges3} and is determined self-consistently as follows: 
firstly, given the local self-energies  $\Sigma^{(i)}(\sigma,i\omega_n)$ obtained from solving the action (\ref{Seff}), the interacting lattice Green's function
follows from the Dyson equation in real-space representation 
\begin{equation}
\mathbf G(\sigma,i\omega_n)^{-1}=\mathbf G_0(\sigma,i\omega_n)^{-1} - \mathbf\Sigma(\sigma,i\omega_n),
\label{realspacedyson}
\end{equation}
where a boldface notation indicates that the quantities are matrices labeled by site indices $i$ and $j$ and
$\omega_n$ are the Matsubara frequencies. 
The non-interacting lattice Green's function is given by  \mbox{$\mathbf G_0(\sigma,i\omega_n)^{-1}=(\mu_{\sigma}+i\omega_n)\mathbf 1 - \mathbf J - \mathbf V,$} where $\mathbf 1$ is the unity matrix.
 The matrix elements $J_{ij}$ are hopping amplitudes for a given lattice structure and $V_{ij}=\delta_{ij}V_i$ represents a spatially varying potential.
Secondly, 
the diagonal elements of the lattice Green's function are identified with the interacting local Green's functions, i.e. 
$G^{(i)}(\sigma,i\omega_n) = G_{ii}(\sigma,i\omega_n)$.
Finally, the Weiss mean-field is obtained from the local Dyson equation 
\begin{equation}
\label{concreteselfcon}
 \mathcal G^{(i)}_0(\sigma,i\omega_n)^{-1} = G^{(i)}(\sigma,i\omega_n)^{-1} + \Sigma^{(i)}(\sigma,i\omega_n),
\end{equation}
which closes the set of the self-consistency equations. 

The most difficult step in this procedure is the solution of the local action (\ref{Seff}). This step is, however, similar to the solution of the local action in a homogeneous DMFT calculation. The difference is that in the present case the Weiss field $\mathcal G^{(i)}_0(\sigma,\tau)$ is obtained via the Real-Space Dyson equation (\ref{realspacedyson}), which incorparates the effect of the spatial inhomogeneity. This implies that for the numerical solution of the local action we can use standard techniques, which have proven to be reliable and efficient. In the present manuscript we use the Numerical Renormalization Group (NRG) at $T=0$ \cite{nrg1, nrg2, nrg3, nrg4, NRG} to solve the single site problems.

In practice the self-consistent solution is obtained iteratively from the initial Weiss mean-fields $\mathcal G^{(i)}_0(\sigma,i\omega_n)$ which are chosen differently for different spin $\sigma$ and lattice sites $i$. Then the solutions with staggered magnetization or phase separation are obtained naturally in contrast to the standard DMFT, where additional sublattice structure has to be added \cite{Georges3}. 

Within R-DMFT significantly larger systems can be investigated than those studied by quantum Monte Carlo \cite{Assaad, Staudt, Rigol}, for which in two and three dimensions only homogeneous data are available. 
The computational effort scales polynomially with the number of lattice sites $N$ within R-DMFT. The application of the real-space Dyson equation requires a sparse matrix inversion for each frequency, which scales as $N^{3/2}$. 
The number of NRG calculations per R-DMFT-run is linear in $N$, but can be kept small due to symmetries. 
Moreover, the solution of the real-space Dyson equation can be parallelized over the frequencies and the NRG-calculations can be parallelized over the lattice sites.

\begin{figure}
\begin{center}
\includegraphics[scale = 0.66, keepaspectratio   ]{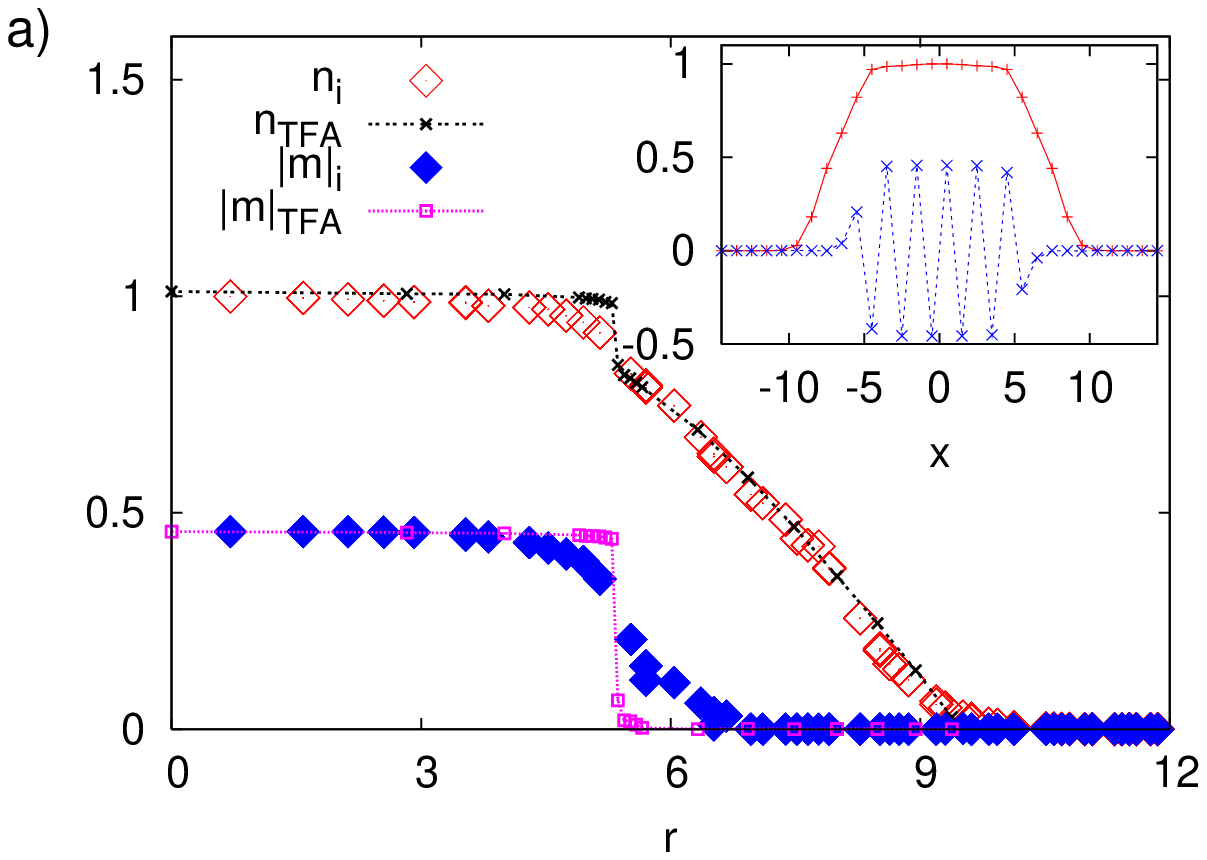}
\includegraphics[scale = 0.66, keepaspectratio   ]{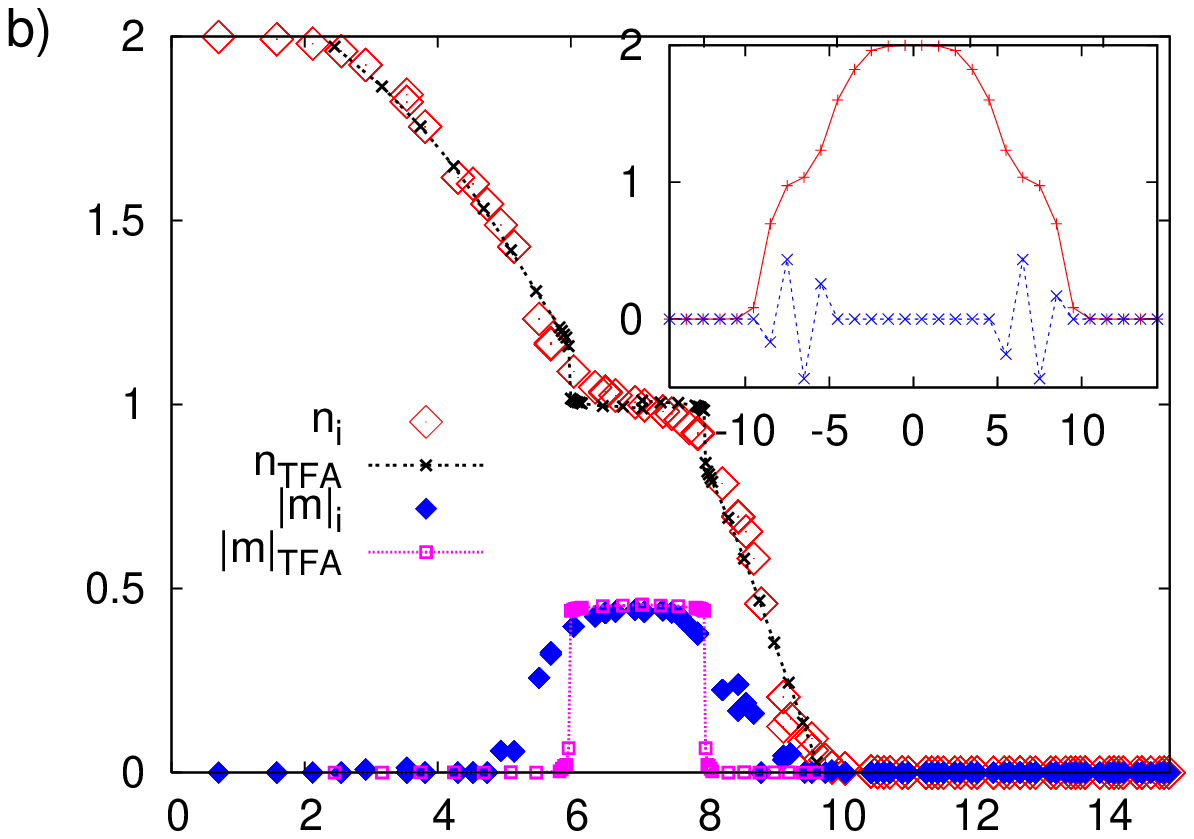}
\end{center}
\vspace{-0.5cm}
\caption{\label{af} 
Particle and spin density profiles determined within the exact solution of R-DMFT and within the Thomas-Fermi approximation (TFA) to R-DMFT.
The main panel shows the local total density $n_{i}$ and the local magnetization $m_i$ as a function of a distance from the trap center.
The inset shows $n_i$ and $m_i$ along the $y=\frac{1}{2}$ line. 
The parameters in the a) and b) panel are the same as in the respective panels of Fig.~\ref{pics}.
}
\end{figure}

\section{Results}
We apply this method to spin-$\frac{1}{2}$ fermions in a two-dimensional square lattice with harmonic confinement. 
In the context of cold atoms a two dimensional system can be realized by applying a highly anisotropic optical lattice, which divides the system into two-dimensional slices. Although not exact, R-DMFT is expected to be a good approximation for the two-dimensional situation at zero temperature, since the derivation of the DMFT equations is controlled by the small parameter $1/z = 1/4$ on the square lattice.

\subsection{Balanced mixture}
First we consider the case of an equal mixture of spin-up and down atoms: $N_\uparrow = N_\downarrow$.
We find that antiferromagnetic order is stable in the presence of the inhomogeneous harmonic potential. 
In Fig.~\ref{pics} we present examples for the spatial dependence of the magnetization at different strengths of the confining potential and the chemical potentials. In the case that the lattice at the center of the trap is half-filled, antiferromagnetism appears in the center of the system  (Fig.~\ref{pics}a). When the particle density in the center of the trap is higher, antiferromagnetic order forms in a ring enclosing a paramagnetic region (Fig.~\ref{pics}b). 
These results are particularly important for ongoing attempts to realize antiferromagnetic states in optical lattices. 
Namely, we predict that the observation of antiferromagnetic order does not critically depend on the number of atoms in the system. 
For sufficiently strong repulsion between the particles, the necessary condition to find antiferromagnetic order is to prepare the system such that the local filling factor approximates or exceeds one in at least part of the system.
We find no evidence for phase separation or a paramagnetic insulating boundary layer for the $N_{\uparrow}=N_{\downarrow} $ case. 

\begin{figure}
\begin{center}
\includegraphics[scale = .66, keepaspectratio]{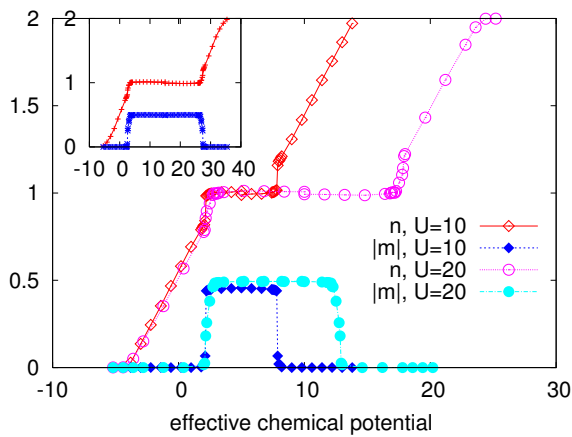}
\end{center}
\vspace{-0.5cm}
\caption{\label{klda}
Total density and staggered magnetization as a function of effective chemical potential obtained within the Thomas-Fermi approximation to R-DMFT for two dimensional square (main panel) and three dimensional cubic lattices (inset). Main panel: $U=10$ (diamonds) and $U=20$ (circles); inset: $U=30$.
}
\end{figure}

The antiferromagnetic ground state of homogeneous fermions described by the Hubbard Hamiltonian (\ref{Hamiltonian}) without trap is stable when the density of particles varies from $n\approx 0.8$ to $1.2$, depending on the interaction value $U$ \cite{Zitzler}. 
On the contrary, in the presence of the external harmonic potential, antiferromagnetic order appears for much lower or higher local total densities. 
Indeed, in Fig.~\ref{af}  we present examples of the local density $n_i$ and the local magnetization $m_i=\frac{1}{2}(n_{i\uparrow}-n_{i\downarrow})$ as a function of distance from the center (main panel) and along a cut through the system (inset) which proves that antiferromagnetic order extends from the center of the trap and disappears only when $n_i\approx 0.5$ in  Fig.~\ref{af}a). 
Similarly, Fig.~\ref{af}b) shows that antiferromagnetic order is stable on a ring when the local density extends between $0.5 \lesssim n_i \lesssim 1.5$. 

We also determine the local density and the local magnetization within the 
Thomas-Fermi approximation (TFA) to R-DMFT,  where the external potential is only included by a spatially varying chemical potential \cite{Byczuk}.  The agreement between the full R-DMFT and TFA results is very good in regions well within or outside the antiferromagnetic domain.
Encouraged by this, Fig.~\ref{klda} shows additional TFA+R-DMFT profiles that can be used to compare R-DMFT with experiments for realistic systems in two and three dimensions. 
However, the staggered magnetization decays abruptly within TFA as compared to the full R-DMFT solution, i.e. the Thomas-Fermi approximation to R-DMFT essentially reproduces results from the standard homogeneous DMFT, cf. Fig.~\ref{klda}. 
The wider stability regime of the antiferromagnetic order found within full R-DMFT is caused by a proximity effect; antiferromagnetic order is induced into parts of the systems where the local densities are too low to stabilize antiferromagnetism in the homogeneous case. On the other hand, the proximity of the paramagnetic state reduces the staggered magnetization when the local density is close to one.

\subsection{Imbalanced mixture}
\begin{figure}
\begin{center}
\includegraphics[scale =1.25]{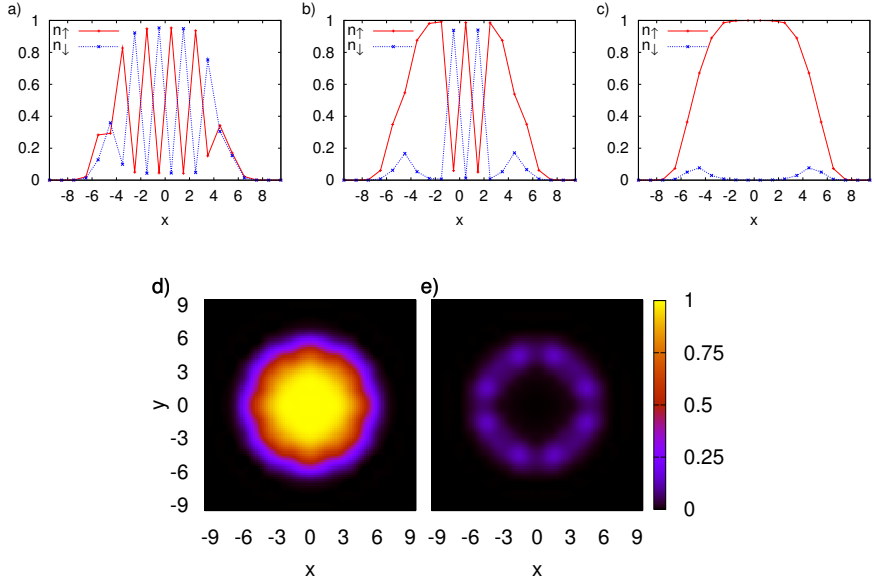}
\vspace{-0.5cm}
\end{center}
\caption{\label{afimb}
Spin resolved particle densities for an imbalanced mixture obtained within R-DMFT for $U=10$.
Panels (a-c) show component densities along the $y=\frac{1}{2}$ line for gradually increasing imbalanced $\Delta\mu=0.3$ (a), $0.75$ (b), $1$ (c).
The two lower panels show the space resolved up- (d) and down- (e) density for $\Delta \mu=1$.
The lattice size is $20\times 20$ and other parameters are: $V=0.2 $, $\bar \mu = 5$.
}
\end{figure}

We now proceed by investigating the imbalanced case \cite{canted}, i.e. $N_{\uparrow}\neq N_{\downarrow}$.  Imbalance between the two spin-components is induced by a nonzero chemical potential difference $\Delta \mu = \mu_\uparrow - \mu_\downarrow$, which corresponds to a magnetic field. In the experimental situation, the density imbalance can be highly tuned and is stable due to the suppression of spin-flip scattering proccesses in cold-atomic gases. 
Representative results are presented in Fig.~\ref{afimb}, where we plot the up- and down-component of the density along a cut through the system. Upon increasing the imbalance parameter $\Delta \mu$, we find suppression of the antiferromagnetic order and emergence of phase separation between the minority and majority species. The phase separation region starts to develop far away from the  center of the trap at small $\Delta \mu$ and gradually spreads toward the center. 
We thus find that the border of the anti-ferromagnetic domain is most sensitive to phase separation. This is indeed reasonable: the energy cost to polarize the antiferromagnetic state is the energy difference between an antiferromagnetic state and a ferromagnetic state. This is of the order $J^2/U$, which small for the large interaction $U$ considered here. The antiferromagnetic order is thereby more unstable for larger distances to the trap-center, because of the vicinity to the paramagnetic region. The energy cost to polarize the paramagnetic regime is higher, because in this case kinetic energy has to be paid, whereas in the anti-ferromagnetic domain the kinetic energy is already quenched because the particles are almost localized. Due to the proximity effect we find that the paramagnetic regime close to the insulating domain also gets phase-separated, which leads to a ring-like structure of the minority species.
 
\begin{figure}
\begin{center}
\includegraphics[scale =0.35]{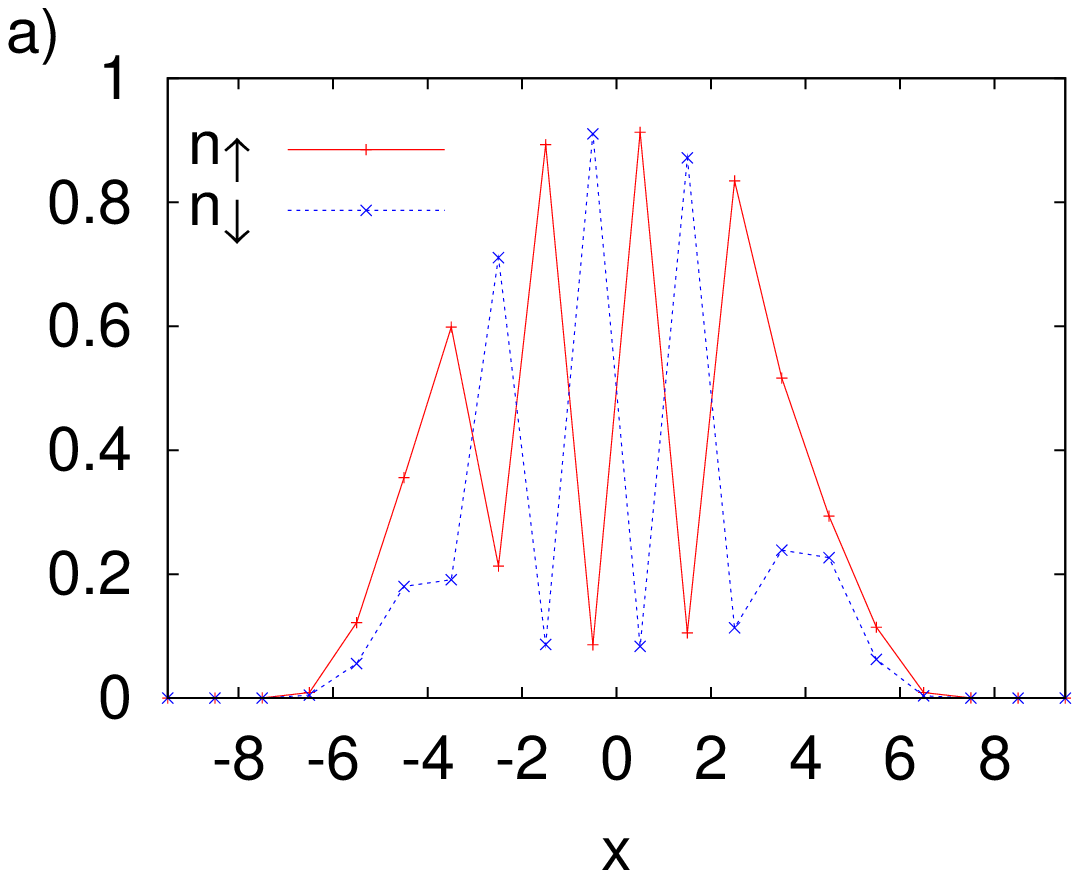}
\includegraphics[scale =0.35]{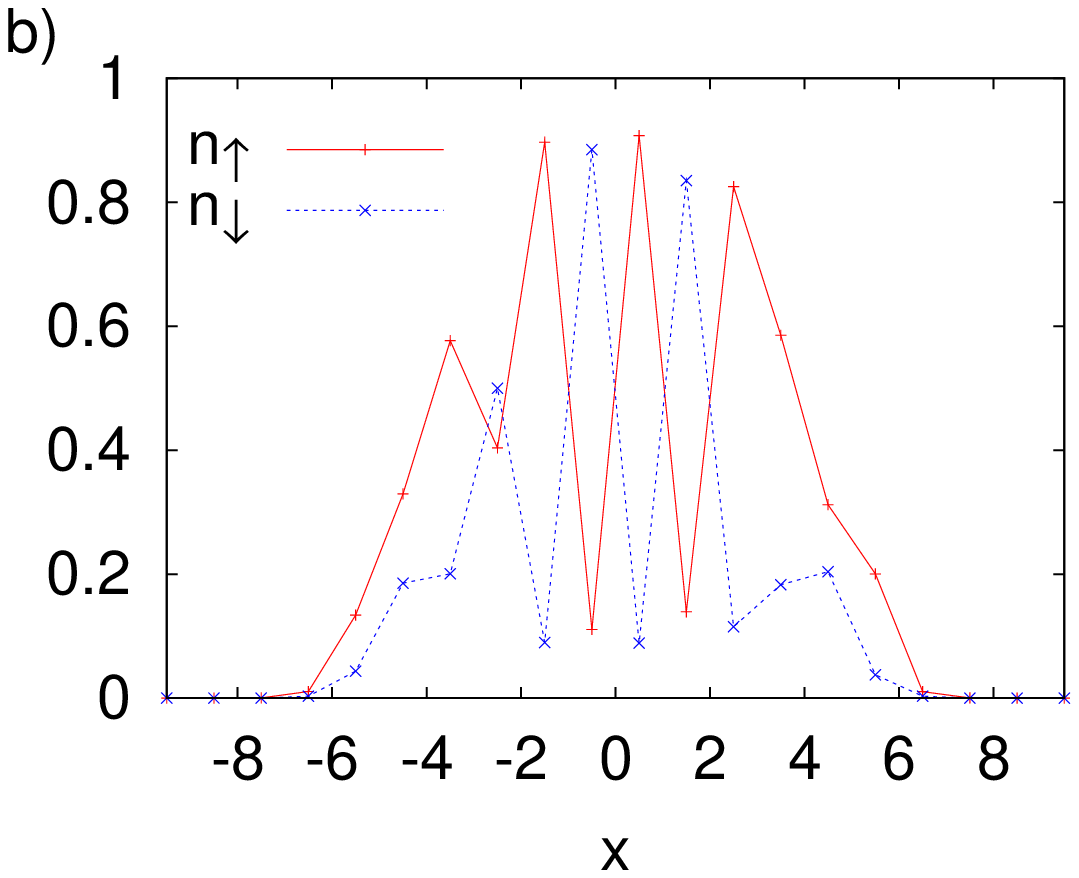}
\includegraphics[scale =0.35]{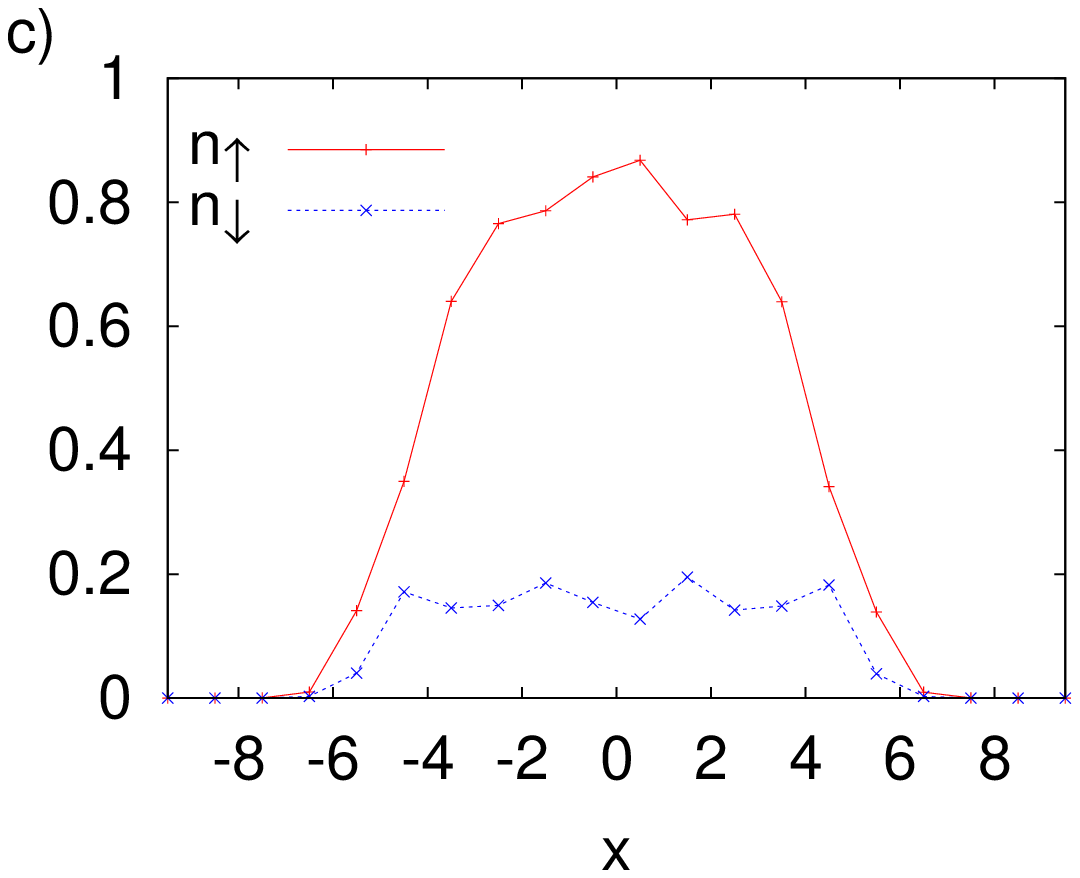}
\end{center}
\caption{\label{afimbU7.5}
Spin resolved particle densities for an imbalanced mixture obtained within R-DMFT for $U=7.5$.
Panels (a-c) show component densities along the $y=\frac{1}{2}$ line for gradually increasing imbalanced $\Delta\mu=0.4$ (a), $0.6$ (b), $8$ (c).
The lattice size is $20\times 20$ and other parameters are: $V=0.2 $, $\bar \mu = 3.75$.
}
\end{figure}

At strong interaction, $U=10$ in the case shown in Fig.~\ref{afimb}, atoms with different spins ultimately tend to occupy different spatial regions to avoid the mutual interaction and the minority species is completely expelled from the trap center. At weaker interaction, however, we found that the imbalanced system still contains interpenetrating atoms with different spins and phase separation does not occur. This is shown in Fig.~\ref{afimbU7.5}, where for $\Delta \mu = 0.8$ the antiferromagnetic order has completely disappeared, but the two spin components are still interpenetrating. The small oscillations in the component densities can be understood as Friedel oscillation due to the  small size of the system. We note that in the case of imbalanced spin-mixtures the agreement between the TFA and the exact solution to R-DMFT is far less good than in the balanced case presented above.

\section{Conclusions}
In conclusion we used the R-DMFT to establish stability of antiferromagnetism for balanced fermionic spin-$\frac{1}{2}$ systems in a trap and the appearance of phase separation for imbalanced mixtures. The antiferromagnetic order predicted here can be observed at low enough temperatures by Fourier-sampling of time-of-flight images 
via Raman pulses \cite{Duan},
by measuring spin correlation functions via local probes \cite{Zhang}, probing noise correlations \cite{Altman, Rom},
polarization spectroscopy \cite{Eckert}, and Bragg scattering \cite{Stenger99}. The effect of spatial inhomogeneity on these probes will be investigated within R-DMFT in future studies. Moreover, the R-DMFT scheme presented here opens up the possibility to study a variety of other strongly correlated systems in inhomogeneous environments.

\ack

We thank I. Bloch, T. Enss, W. Ketterle, D. Vollhardt, and M. Zwierlein for useful discussions. This 
work was supported by the German Science Foundation 
DFG via grant HO 2407/2-1 and the Collaborative Research Centers SFB-TRR 49  and SFB 484.


\section*{References}

\newcommand{\PRB}{Phys.\ Rev.\ B}
 \newcommand{\PRA}{Phys.\ Rev.\ A}


\begin{thebibliography}{99}
\bibitem{NRG}
We use for all of the R-DMFT calculations the NRG parameters $\Lambda = 2.0$, $N_{\rm iter} = 60$ and $600$ kept states. For the TFA-data $1000$ states were kept.
\bibitem{canted} The elements of the Green's function that are off-diagonal in spin-space are assumed to be zero. Therefore we cannot account for canted antiferromagnetism.
\end{thebibliography}
\end{document}